\documentstyle[12pt,amssymb]{article}

\begin{document}
\newcommand{\di}{\frac{d}{di}}
\baselineskip16pt

\title{\vspace*{-10mm}
\sc Axiomatic Holonomy Maps and Generalized Yang-Mills Moduli Space}
\author{\sc Piotr M. Hajac\thanks{
http://info.fuw.edu.pl/KMMF/ludzie\underline{~~}ang.html 
(E-mail: pmh@fuw.edu.pl)}\\
\normalsize Department of Mathematics, University of California, 
Berkeley, CA 94720, USA \vspace*{-0mm}}
\date{1993}
\maketitle
\begin{abstract}
{\parindent0pt
This article is a follow-up of \cite{b}. 
Its main goal is to provide an alternative proof of
 this part of the reconstruction theorem which concerns the 
existence of a connection. 
$\mbox{A construction of connection 1-form is presented.}$ The formula
expressing the local coefficients of connection in terms of the holonomy map is 
obtained as an immediate consequence of that construction. 
$\!\mbox{Thus the derived 
formula coincides with that used in $\!$\cite{csta}.}$
The 
reconstruction and representation theorems form a generalization of the fact that
the pointed configuration space of the classical Yang-Mills theory is equivalent
to the set of all holonomy maps. The point of this generalization is that there is 
a one-to-one correspondence not only between the holonomy maps and the orbits in the
space of connections, but also between all maps $\Omega M\rightarrow G$ fulfilling
some axioms and all possible equivalence classes of $P(M,G)$ bundles with 
connection, where the equivalence relation is defined by bundle isomorphism in a
natural way.}
\end{abstract}

\section{Introduction}

Apart from purely mathematical motivation, which is to provide an alternative
description for the moduli space of bundles with 
connections, there is also a physical incentive
to study gauge theory by means of holonomy maps. The Aharonov-Bohm experiment 
indicates that neither a Yang-Mills potential (i.e. 
$\!$connection) nor a field strength
(i.e. $\!$curvature) is an object corresponding directly to a given physical
 situation
-- connection possesses redundant degrees of freedom and curvature does not 
fully describe the situation in the case of a non-simply connected spacetime
\cite{ct86}. 
On the other hand, holonomy map can be measured and it provides more information
than curvature \cite{a83,a86}.
The reconstruction theorem gives the mathematical ground for treating 
holonomy as the primary, and potential and field strength (including the total
space of the principal bundle on which they are defined) as the secondary, derived,
objects of the theory \cite{a83,b,csta}.

The holonomy, or
loop space, approach was applied to derive
the equations of motion of the nonabelian monopole \cite{cstb,ct}.
Furthermore, there is an extension of the holonomy formalism used in Yang-Mills 
theory which is applied in general relativity. That extension, including physical
motivations standing behind it, is described in \cite{b}. With some
negligible exceptions, the notations and conventions used throughout this paper are
consistent with those of \cite{b}. 
  The next two paragraphs are devoted to establishing 
notation, recalling the reconstruction and representation theorems and the bundle
construction. In paragraph 4, an alternative proof of the existence of the holonomy
reconstructed connection is presented. An advantage of this proof is that it allows
to write an explicit formula for the action of connection 1-form [see (5.1)] and,
as a consequence, gives a geometrical interpretation for the local expression of
the holonomy defined Yang-Mills potential \cite{csta}.

\section{Notations and Conventions}

\begin{description}

\item[$\widetilde{\psi}$] family of paths: $\widetilde{\psi} : U\rightarrow PM,\;
\widetilde{\psi}[u](i):=\psi (u,i)$, where $\psi : U\times [0,1]\rightarrow
 M$ is 
the
associated map of $\widetilde{\psi}$; $\widetilde{\psi}$ is called smooth iff 
$\psi$ is
continuous and smooth on the subintervals $U\times [i_{n},i_{n+1}]$ for 
$i_{0}=0<i_{1}<...<i_{k}<i_{k+1}=1,\; n\in \{0,...,k\}$

\item[$a^{-1}$] diffeomorphism $: \pi ^{-1}(\pi (a))\rightarrow G$ defined by
$a^{-1}b=g\Leftrightarrow b=ag$, where $\pi : P\rightarrow M$ is the projection of
the principal fiber bundle $P(M,G)$
\item[${\cal F}$] the set of all triples 
$\mbox{\boldmath $($}P(M,G)\mbox{\boldmath $,$}\omega 
\mbox{\boldmath $,$}b\mbox{\boldmath $)$}$, where $M$
and $G$ are fixed and $\omega\!\in\! C(P),\, b\!\in\!\pi ^{-1}(*)$
\item[${\cal F}/S$] generalized configuration space\footnote{
If instead of fixing $M$ and $G$ we fix the entire bundle $P(M,G)$ and the point 
$b\in \pi ^{-1}(*)$, then ${\cal F}$ will collapse to $C(P)$ and ${\cal F}/S$ will
 become the pointed configuration space of Yang Mills theory, i.e. $\!
C(P)/GA_{*}(P)$, where
$GA_{*}(P):=\{ f\in Diff(P)|\forall g\in G:\; f\circ R_{g}=R_{g}\circ f,\; \pi\circ
f=\pi ,\; f(b)=b\}$.}; $S$ is an equivalence
relation defined by: 

$\mbox{\boldmath $($}P(M,G)\mbox{\boldmath $,$}\omega 
\mbox{\boldmath $,$}b\mbox{\boldmath $)$}\;
S \;
\mbox{\boldmath $($}P^{\prime}(M,G)\mbox{\boldmath $,$}
\omega ^{\prime}\mbox{\boldmath $,$}b^{\prime}\mbox{\boldmath $)$}
\;\mbox{\large\boldmath $\Longleftrightarrow\exists $}f\in Diff(P,P^{\prime
})\mbox{\large\boldmath $:$}\;\;
\pi ^{\prime}\circ f=\pi \mbox{\large\boldmath $,$}$

$\forall g\in G:\;
f\circ R_{g}=R^{\prime}_{g}\circ f\mbox{\large
\boldmath $,$} \;\;\; (f^{-1})^{*}\omega =\omega
^{\prime}\mbox{\large\boldmath $,$}\;\;\; f(b)=b^{\prime}$

\item[$\mbox{thin loop}$] For precise definitions see p.1180 and 
p.1205 in \cite{b}, \cite{k} and 6.1 in \cite{m}.

\item[${\cal H}$] the space of axiomatically defined holonomy maps: 
\[
{\cal H}:=\{ H:\Omega M\rightarrow G\; |\;\mbox{$H$ satisfies axioms 1,2,3}\}
\]

\item[Axiom 1] $\forall\alpha ,\beta\in\Omega M:\; H(\alpha\circ\beta
)=H(\beta )H(\alpha )\;$ (The loop $\alpha\circ\beta$ is the one obtained
by first going around the loop $\beta$ and then around $\alpha$.)

\item[Axiom 2] $H$(any thin loop) $=e\;$ (Note that this axiom
makes any two loops which differ only by their parametrization indistinguishable for
any $H\in {\cal H}$.)

\item[Axiom 3] For every smooth family of loops 
$\widetilde{\Omega}:U\rightarrow\Omega M$, $U$ an open subset of ${\Bbb R}^{n}$,
$n\in {\Bbb N}$, $H\circ\widetilde{\Omega}:U\rightarrow G$ is smooth.

\end{description}
{\parindent=0pt
\section{The Reconstruction and Representation Theorems
and the Bundle Construction}

With notation assumed as above, the representation and reconstruction theorems are
equivalent to:
\newtheorem{theorem}{Theorem}[section]
\begin{theorem}
Let $M$ be a Hausdorff connected manifold and $H^{\omega}_{b}$ the holonomy map
w.r.t. $\!$the point $b$ and connection $\omega $ (i.e. $\! H^{\omega}_{b}:\Omega
M\ni\alpha\mapsto\check{\alpha }(1)^{-1}\check{\alpha}(0)$, $\pi\circ\check{
\alpha}=\alpha$, $\check{\alpha}(0)=b$, $\omega\frac{d}{di}\check{\alpha}(i)=0$). 
The map
\[
\mbox{hol}:{\cal F}/S\ni\mbox{\large\boldmath $[$}\mbox{\boldmath $($}P(M,G)\mbox{
\boldmath $,$}\;\omega\,\mbox{\boldmath $,$}\; b
\mbox{\boldmath $)$}\mbox{\large\boldmath $]$}\longmapsto H^{\omega}_{b}\in {\cal H}
\]
is well defined and bijective.
\end{theorem}}
The representation part of Theorem 3.1 asserts that the map $hol$ exists and is 
injective, whereas the reconstruction part claims that $hol$ is surjective.
The following are the geometric structures used to prove the theorem:
$E:=(PM\times G)/T$ (total space), where 
$(p,g)\; T\; (p^{\prime},g^{\prime})\stackrel{\rm def}{\Longleftrightarrow} 
\mbox{\boldmath $($}\; p(1)=p^{\prime}(1)$ and 
$g^{\prime}=H(p^{-1}\circ p^{\prime})g\;\mbox{\boldmath $)$}$, $\{ p,g\}=\{ p^{
\prime},g^{\prime}\}\in E$; $R_{g}\{ p,h\} :=\{ p,hg\}$ (right action of $G$);
$\pi \{ p,h\} :=p(1)$ (projection); $\widehat{p}(i):=\mbox{
\boldmath $\{$}K(p,i)\circ q\;\mbox{\bf ,}\; g\mbox{\boldmath $\}$}$
(lifting of paths, to become the horizontal lifting),
where $q\in PM,\; q(1)=p(0)$ and $K(p,i)$ is the contraction of $p$, i.e. 
$K(p,i)(j):=p(ij)$; local trivializations:
\[
C_{\psi}:U\times G\stackrel{(\widetilde{\psi},id)}
{\longrightarrow}PM\times G\longrightarrow\pi ^{-1}(U)\subseteq (PM\times G)/T
\]\[
C_{\psi}^{-1}\{ p,g\} =C_{\psi}^{
-1}\mbox{\large\boldmath $\{$}\widetilde{\psi}[p(1)]\;\mbox{\large\bf ,}\;
H\mbox{\bf (}p^{-1}\circ\widetilde{\psi}[p(1)]\mbox{\bf )}g\mbox{\large\boldmath
$\}$}
=\mbox{\large\boldmath $($}\; p(1)\;\mbox{\large\bf ,}\;
H\mbox{\bf (}p^{-1}\circ\widetilde{\psi}[p(1)]\mbox{\bf )}g\mbox{\large
\boldmath $)$},
\]
 where $U$ is a contractible open subset of $M$ and $\widetilde{
\psi}$ is a smooth family of paths $\mbox{$\widetilde{\psi}:U\rightarrow PM$}$ 
 having the
property $\forall u\in U:\; \widetilde{\psi}[u](1)=u$.

\section{Reconstruction of Connection}

\newtheorem{proposition}{Proposition}[section]
\begin{proposition}
Let $\Gamma _{\{ p,g\} }$ denote the vector subspace of $T_{\{ p,g\} }E$ generated
by vectors of the form $\frac{d}{di}\mbox{\boldmath $\{$}K(r,i)\mbox{\boldmath $,$}
\, h\mbox{\boldmath $\}$}\mid _{i=j}$, where 
$h:=H\mbox{\boldmath $($}p^{-1}\circ K(r,j)\mbox{\boldmath $)$}g$. Then the 
distribution $\Gamma :E\ni a\mapsto 
\Gamma _{a}$ is a smooth connection.
\end{proposition}

{\bf Proof:}

{\parindent=0pt 
{\bf a)} $\mbox{\large\boldmath $\forall $}\{ p,g\}\in E\mbox{\large\boldmath $:$}
\;\; \Gamma _{\{ p,g\} }\oplus
Ker(\pi _{*}:T_{\{ p,g\} }E\rightarrow T_{p(1)}M)=T_{\{ p,g\} }E$ \\
 That follows from the following lemmas:
\smallskip

\newtheorem{lem}{Lemma}
\begin{lem}
\[
\mbox{\large\boldmath $\forall $}
 a\in E,\; X_{a}\in T_{a}E\; 
\mbox{\large\boldmath $\exists $}p\in PM,\; k\in C^{\infty}([0,1],G)\mbox{\large
\boldmath $:$}
\]\[
X_{a}=\frac{d}{di}\{ K(p,i),k(j)\}\mid _{i=j}+\,\frac{d}{di}\{ K(p,j),k(i)\}\mid 
_{i=j}
\]
\end{lem}}

{\sl Proof:}

{\parindent=0pt
It is clear that for an arbitrary $X_{a}$ there exist smooth curves p and g such
that
$X_{a}=C_{\psi *}\frac{d}{di}\mbox{\bf (}p(i)\mbox{\bf ,}\, g(i)\mbox{\bf )}\mid
_{i=j}$. If we put $k(i)=H\mbox{\boldmath $($}\widetilde{\psi}[p(i)]^{-1}\circ
K(p,i)\mbox{\boldmath $)$}g(i)$, we have:
\[
X_{a}=\frac{d}{di}\mbox{\boldmath $\{$}\widetilde{\psi}[p(i)]\;
\mbox{\boldmath $,$}\; g(i)\mbox{\boldmath $\}$}\mid _{i=j}\;
=\frac{d}{di}\mbox{\boldmath $\{$}K(p,i)\;\mbox{\boldmath $,$}\;
k(i)\mbox{\boldmath $\}$}\mid _{i=j}\; =\phantom{.kfvmnamvaienmrgvonrgooiokeknvkndk
fnklamnvklfmnlkklklfv}
\]\[
=C_{\psi *}\frac{d}{di}C_{\psi}^{-1}\mbox{\boldmath $\{$}K(p,i)\;
\mbox{\boldmath $,$}\; k(i)\mbox{\boldmath $\}$}\mid _{i=j}\;
 =C_{\psi *}\frac{d}{di}\mbox{\large\boldmath $($}p(i)\;\mbox{\large\boldmath $,$}
\; H\mbox{\boldmath $($}K(p,i)^{-1}\circ\widetilde{
\psi}[p(i)]\mbox{\boldmath $)$}k(i)\mbox{\large\boldmath $)$}\mid _{i=j}\; 
=\phantom{kgbmfkgmblkfgmbl'kfgmb'lkfgmb'lkfgmb'lkgmblkmgkggfmk.}
\]\[
=C_{\psi *}\frac{d}{di}\mbox{\large\boldmath $($}p(i)\mbox{\large\boldmath $,$
}\! H\mbox{\boldmath $($}K(p,i)^{-1}\circ\widetilde{
\psi}[p(i)]\mbox{\boldmath $)$}k(j)\mbox{\large\boldmath $)$}\!\!\mid _{i=j}\! +C_{
\psi *}\frac{d}{di}\mbox{\large\boldmath $($}p(j)\mbox{\large\boldmath $,$
}\! H\mbox{\boldmath $($}K(p,j)^{-1}\circ\widetilde{
\psi}[p(j)]\mbox{\boldmath $)$}k(i)\mbox{\large\boldmath $)$}\!\!
\mid _{i=j}=\phantom{
kmgfblknfgsbjnggbjnfgbjnfgbjmfgkbmflgkbnmlkgbkgbnbknkgfk.}
\]\[
=\frac{d}{di}\mbox{\large\boldmath $\{$}\widetilde{\psi}[p(i)]\mbox{\large
\boldmath $,$}\, H\mbox{\boldmath $($}K(p,i)^{-1}\circ\widetilde{
\psi}[p(i)]\mbox{\boldmath $)$}k(j)\mbox{\large\boldmath $\}$}\!\!\mid
_{i=j}+\;\frac{d}{di}\mbox{\large\boldmath $\{$}\widetilde{\psi}[p(j)]\mbox{\large
\boldmath $,$}\, H\mbox{\boldmath $($}K(p,j)^{-1}\circ\widetilde{
\psi}[p(j)]\mbox{\boldmath $)$}k(i)\mbox{\large\boldmath $\}$}\!\!\mid 
_{i=j}=\phantom{kfgbmklfgmblkfgnmbjlfgsnbjfgnbjfgnbjgnbjnsgbjn.}
\]\[
=\frac{d}{di}\mbox{\boldmath $\{$}K(p,i)\;\mbox{\boldmath $,$}\;
k(j)\mbox{\boldmath $\}$}\mid _{i=j}+\,\frac{d}{di}\mbox{\boldmath $\{$}K(p,j)\;
\mbox{\boldmath $,$}\; k(i)\mbox{\boldmath $\}$}\mid _{i=j}\;\;\Box\phantom{jfgnjn
skjfgbnksjbgsdbvsdhkfbvsdkfbvkldfbvskdfjbgdijfbvdfbvsdfkbvsdkdfhjlbnsfjkgns}
\]

\newtheorem{lemm}[lem]{Lemma}
\begin{lemm}
$\mbox{\boldmath $\forall $} a\in E\mbox{\boldmath $:$}\;\; \pi _{*}:\Gamma _{a}
\rightarrow T_{\pi (a)}M$
is an isomorphism.
\end{lemm}
}

{\sl Proof:}

{\parindent=0pt 
\[
\pi _{*}\frac{d}{di}\mbox{\boldmath $\{$}K(p,i)\;\mbox{\bf ,}\;
 h\mbox{\boldmath $\}$}\mid _{i=j}\; =\frac{d}{di}p(i)\mid _{i=j}\;
\mbox{\boldmath $\Rightarrow $}\;\pi _{*}\; \mbox{is onto}
\]
Furthermore, 
\[
\pi _{*}\frac{d}{di}\mbox{\boldmath $\{$}K(p,i)\;\mbox{\bf ,}\; 
h\mbox{\boldmath $\}$}\mid _{i=j}\;
 =0\;\mbox{\boldmath $\Rightarrow $
}\;\frac{d}{di}p(i)\mid
_{i=j}\; =0\;\mbox{\boldmath $\Rightarrow $}
\]\[
\frac{d}{di}\mbox{\boldmath $\{$}K(p,i)\;\mbox{\bf ,}\;
h\mbox{\boldmath $\}$}\mid _{i=j}\; =C_{\psi *}\frac{d}{di}\mbox{\large\boldmath $($
}p(i)\;\mbox{\large\bf ,}\;
H\mbox{\boldmath $($}K(p,i)^{-1}\circ\widetilde{\psi}[p(i)]\mbox{\bf )}h\;
\mbox{\large\boldmath $)$}\mid
_{i=j}\;
=\phantom{w.wh.=x.}
\]\[
=C_{\psi *}\mbox{\large\boldmath $($}\; 0\;\mbox{\large\bf ,}\;
\frac{d}{di}\mbox{\bf (}H\circ\widetilde{\Omega
}\mbox{\bf )(}p(i)\mbox{\bf )}\mid _{i=j}\mbox{\large\boldmath $)$}=C_{\psi *
}\mbox{\large\boldmath $($}\; 0\;\mbox{\large\bf ,}\;
\partial_{\mu}\mbox{\bf (}H\circ\widetilde{\Omega
}\mbox{\bf )}\frac{d}{di}p^{\mu }(i)\mid _{i=j}\mbox{\large\boldmath $)$}=0\phantom{
w.w.h.}
\]
where $\widetilde{\Omega }[p(i)]:=K(p,i)^{-1}\circ\widetilde{\psi}[p(i)]$. 
$\Box$

{\bf b)} $R_{g *}\Gamma _{a}=\Gamma _{ag}$ (trivial)

{\bf c)} $\Gamma $ is smooth

Since smoothness is a local property, we are interested only in some open 
neighborhood of an arbitrarily chosen point $m\in M$ and therefore can treat that 
neighborhood as ${\Bbb R}^{n}$, $\mbox{$n:=dim M$},\; m=0$.
By Lemma 2, it is enough to prove that $\forall\mu\in\{ 1,...,n\} $ the lifting
$\widehat{\partial _{\mu }}$ of the coordinate vector field $\partial _{\mu }$ is
smooth. Let $\widetilde{\psi}$ and
$\{\widetilde{\varphi _{\mu}}\} _{\mu\in\{1,...,n\}}$ be smooth families of paths
and $\{\widetilde{\Omega _{\mu}}\} _{\mu\in\{ 1,...,n\} }$ smooth families of loops
defined by:

\[
\psi (x,i)=\left\{\begin{array}{ll}
q(2i) & \mbox{for $0\leq i\leq\frac{1}{2},\; q\in PM,\; q(1)=0$} \\
(2i-1)x & \mbox{for $\frac{1}{2}\leq i\leq 1,\; x:=(x_{1},...,x_{n})\in {\Bbb R}^{n}
$}\end{array}\right.
\]
\[
\varphi _{\mu}(x,i)=\left\{\begin{array}{ll}
q(4i)\;\mbox{for $0\leq i\leq\frac{1}{4}$, $q$ and $x$ the same as above}\\
(4i-1)(x_{1},...,x_{\mu}-\frac{1}{4},...,x_{n})\;\;\;\;\,\mbox{for}\;\;\;\;\,
\frac{1}{4}\leq i\leq\frac{1}{2} \\
(x_{1},...,x_{\mu}-\frac{3}{4}+i,...,x_{n
})\phantom{,x...}\;\;\;\;\,\mbox{for}\;\;\;\;\,
\frac{1}{2}\leq i\leq 1
\end{array}\right.
\]

\[
\widetilde{\Omega _{\mu}}[(x,i)]=\widetilde{\psi}[x]^{-1}\circ T^{\mu}_{x,i}\circ
\widetilde{\psi}\mbox{\boldmath
$[$}\widetilde{\varphi _{\mu}}[x](i)\mbox{\boldmath $]$},\;\; (x,i)\in {\Bbb R}^{n
}\times (\frac{1}{2},1),\phantom{jhu;othurgeuohuhuht3urpv3uprhtviuhrituhvhiuwbviuwwrjwrhbnvuwhrviuwbvbwv}
\]
\[
T^{\mu}_{x,i}(j):=\widetilde{\varphi _{\mu}
}[x](i)+j\mbox{\boldmath $($}x-\widetilde{\varphi _{\mu}
}[x](i)\mbox{\boldmath $)$}=(1-j)(x_{1},...,x_{
\mu}-\frac{3}{4}+i,...,x_{n})+jx\phantom{jhdfbvjahbfvjhabvhfhvhjvhabvjhafbvjhabvja}
\]

Clearly, $\widetilde{\psi}[x](1)=x$ and $\frac{d}{di}\widetilde{\varphi _{\mu
}}[x](i)\mid _{i=\frac{3}{4}}=\partial _{\mu}(x)$. Since the differential structure 
on $E$ is given by its local trivializations and $C_{\psi}$ is a local
 trivialization of $E$, it suffices to show that
$C_{\psi *}^{-1}\widehat{\partial _{\mu}}$ is a smooth vector field on
 ${\Bbb R}^{n}\times G$:
\[
(C_{\psi *}^{-1}\widehat{\partial _{\mu}})(x,g)=C_{\psi *}^{-1}\widehat{
\partial _{\mu}}\mbox{\boldmath $($}C_{\psi}(x,g)\mbox{\boldmath $)$}=C_{
\psi *}^{-1}\widehat{\partial _{\mu}}\mbox{\boldmath $($}\{ \widetilde{\psi
}[x],g\}\mbox{\boldmath $)$}=\phantom{iyfbbfhghgjhjhvgvghvhgvghhgghghghghhgeyqfbhbhbhbvhbhbvjhbvhvhbbhebvyieubbyyg}
\]\[
=C_{\psi *}^{-1}\frac{d}{di}\mbox{\large\boldmath $\{$} K(\widetilde{\varphi _{\mu
}}[x],i)\;\mbox{\large\bf ,}\; 
H\mbox{\boldmath $($}\widetilde{\psi}[x]^{-1}\circ K(\widetilde{
\varphi _{\mu}}[x],\frac{3}{4})\mbox{\boldmath $)$}g
\mbox{\large\boldmath $\}$}\mid _{i=\frac{3}{4}}=\phantom{iyutfgyuyyugyugyugyugyuggggggggggggggggggggggggggggggggggggggfbbfeyqebvyieubbyyg}
\]\[
=\frac{d}{di}\mbox{\Large\boldmath $($}\widetilde{\varphi _{\mu}
}[x](i)\;\mbox{\Large\bf ,}\; H\mbox{\large\boldmath $($}K(\widetilde{
\varphi _{\mu}}[x],i)^{-1}\circ\widetilde{\psi}\mbox{\boldmath $[$}\widetilde{
\varphi _{\mu}}[x](i)\mbox{\boldmath $]$}\mbox{\large\boldmath $)$}H
\mbox{\large\boldmath $($}\widetilde{\psi}[x]^{-1}\circ K(\widetilde{
\varphi _{\mu}}[x],\frac{3}{4})\mbox{\large\boldmath $)$}g
\mbox{\Large\boldmath $)$}\mid _{i=\frac{3}{4}}=\phantom{.kujfbgvuibfiubau;jggjjggjjjjsjjsklkdijihgbhvbjhvgjvgvgvjghvjhvjhgjhhjghjhjhjgghjghjdil}
\]\[
=\mbox{\boldmath $($}\;\partial _{\mu}(x)\;\mbox{\boldmath $,$}\; 
\frac{\partial}{\partial i}(H\circ\widetilde{\Omega _{\mu}})(x,i)g\mid 
_{i=\frac{3}{4}}\mbox{\boldmath $)$}\phantom{.kujfbgvuibfiubau;jggjjggjjjjsjjsklkd
ijihgbhvbjhvgjvgvgvjghvjhvjhgjhhjghjhjhjgghjghjdil}
\]
Now, the smoothness of $C_{\psi *}^{-1}\widehat{\partial _{\mu}}$ follows from
the smoothness of $H\circ\widetilde{\Omega _{\mu}}$. $\rule{7pt}{7pt}$}

\section{Connection 1-form}

Let $P_{V}$ denote the projection on vertical subspaces and $X_{a}=\frac{d}{di}\{
\widetilde{\chi}[i],g(i)\}\mid _{i=j}$ be an arbitrary vector tangent to $E$. 
The action of connection 1-form on $X_{a}$ can be described as follows:
\[
\omega _{a}:T_{a}E\ni X_{a}\longmapsto (a^{-1})_{*}P_{V}X_{a}\in G^{\prime}
\]
By Lemma 1, we have
\[
X_{a}=\frac{d}{di}\mbox{\boldmath $\{$}K(p,i)\;\mbox{\boldmath $,$}\;
k(j)\mbox{\boldmath $\}$}\mid _{i=j}+\,\frac{d}{di}\mbox{\boldmath $\{$}K(p,j)\;
\mbox{\boldmath $,$}\; k(i)\mbox{\boldmath $\}$}\mid _{i=j},
\]
where $p(i):=\widetilde{\chi}[i](1),\; k(i):=H\mbox{\boldmath $($}\widetilde{\chi
}[i]^{-1}\circ K(p,i)\mbox{\boldmath $)$}g(i)$.
\vspace{3.5mm}

{\parindent=0pt Hence}
\[
\omega _{a}X_{a}=\mbox{\large\boldmath $($}\mbox{\boldmath $\{$}K(p,j)\;
\mbox{\boldmath $,$}\; k(j)\mbox{\boldmath $\}$}^{-1}\mbox{\large\boldmath $)$}_{
*}\frac{d}{di}\mbox{\boldmath $\{$}K(p,j)\;
\mbox{\boldmath $,$}\; k(i)\mbox{\boldmath $\}$}\mid _{i=j}=\phantom{uisdfviubvbbhib
kufhdvuiahfviuakjgbksjbkjsbisrgbvisurhbsurbisurbgnusnbsuhtburshbuhtbuthbuthbh}
\]\[
=\frac{d}{di}
\mbox{\boldmath $\{$}K(p,j)\;
\mbox{\boldmath $,$}\; k(j)\mbox{\boldmath $\}$}^{-1}
\mbox{\boldmath $\{$}K(p,j)\;
\mbox{\boldmath $,$}\; k(i)\mbox{\boldmath $\}$}\mid _{i=j}=
\frac{d}{di}k(j)^{-1}k(i)\mid _{i=j}=\phantom{idfbvildfbvifdgbviajjjjibojbohjbothgbouhtgbouthgbouthuwsthgbouthbothbfdgviaugvagfirivgb}
\]\[
=\frac{d}{di}\mbox{\large\boldmath $($}H
\mbox{\boldmath $($}\widetilde{\chi}[j]^{-1}\circ K(p,j)\mbox{\boldmath $)$}g(j)
\mbox{\large\boldmath $)$}^{-1}H\mbox{\boldmath $($}\widetilde{\chi
}[i]^{-1}\circ K(p,i)\mbox{\boldmath $)$}g(i)\mid _{i=j}=\phantom{jkghbjhgbjshgb
jgnbjkshnbjshbviusrhbkjsgfhbisghbsfigubiuljnbuhbuishbiushbiusbdbvissgbi}
\]\[
=g(j)^{-1}\frac{d}{di}H\mbox{\boldmath $($}K(p,j)^{-1}\circ\widetilde{\chi
}[j]\mbox{\boldmath $)$}H\mbox{\boldmath $($}\widetilde{\chi
}[i]^{-1}\circ K(p,i)\mbox{\boldmath $)$}g(i)\mid _{i=j}=\phantom{igygyigiugygyggy
ygjhgjlhguygufutfvutygofvuyouyguyguugutyguyyguygyuhuubhshbiuhgbvithvbig}
\]\[
=g(j)^{-1}\frac{d}{di}H\mbox{\boldmath $($}\widetilde{\chi
}[i]^{-1}\circ K(p,i)\circ K(p,j)^{-1}\circ\widetilde{\chi
}[j]\mbox{\boldmath $)$}g(i)\mid _{i=j}\hspace{56 mm}\mbox{(5.1)}
\]

Note that $K(p,i)\circ K(p,j)^{-1}$ is a curve beginning at $p(j)$ and ending at
$p(i)$ whose image is contained in the image of $p$.


To compute the local coefficients of $\omega $ we must pick up a local
section of $E$. The natural choice seems to be:
\[
\sigma :U\ni x\longmapsto C_{\psi}(x,e)\in\pi ^{-1}(U),
\]
where $C_{\psi}$ is a local trivialization over $U$. Similarly as in the proof of
the smoothness of $\Gamma $, we are interested only in some open subset of $M$
which is small enough to be diffeomorphic to ${\Bbb R}^{n}$.
 Therefore, in what
follows, $U$ will be identified with ${\Bbb R}^{n}$. Putting $T_{y,x}(i)=x+i(y-x)$
and $T_{y_{\mu},x}(i)=x+i(0,...,y_{\mu}-x_{\mu},...,0)$ and using (5.1) we get:
\[
A_{\mu}(x):=(\sigma ^{*}\omega )_{x}\partial _{\mu}(x)=\omega _{\sigma (x)}\sigma
_{*}\partial _{\mu}(x)=\omega _{\sigma (x)}\sigma _{*}\frac{d}{dy_{\mu}}T_{y_{\mu
},x}(1)\mid _{y_{\mu}=x_{\mu}}=\phantom{iuhvuibviuygfbviyuagoiyuvhvvchgcghcc
uiygiylgigiyuyfuytfutffuytr}
\]\[
=\omega _{\sigma (x)}\frac{d}{dy_{\mu}}\{\widetilde{\psi}[T_{y_{\mu
},x}(1)]\; ,\; e\}\mid _{y_{\mu}=x_{\mu}}=\frac{d}{dy_{\mu}}H
\mbox{\boldmath $($}\widetilde{\psi}[T_{y_{\mu},x}(1)]^{-1}\circ
T_{y_{\mu},x}\circ\widetilde{\psi}[T_{x_{\mu},x}(1)]\mbox{\boldmath $)$}
\mid _{y_{\mu}=x_{\mu}}=\phantom{iuhvuibviuygfbviyuagoiyuvr}
\]\[
=\frac{\partial}{\partial y_{\mu}}H(\widetilde{\psi}[y]^{-1}\circ
T_{y,x}\circ\widetilde{\psi}[x])\mid _{y=x}\hspace{99 mm} (5.2)
\]

Formula (5.2) coincides with formula (5.26) in \cite{csta},
where it is taken to be the definition of the gauge potential $A_{\mu}$.

\section{Example}
Let $M={\Bbb R}^{2},\; *=0,\; G={\Bbb R}_{*}$ and $H(\beta )=exp\int_{\beta}ydx,\;
\beta\in\Omega{\Bbb R}^{2}$. Clearly, the reconstructed bundle $E$ is
trivial and can be identified with ${\Bbb R}^{2}\times{\Bbb R}_{*}$ via the bundle
isomorphism $C_{\psi}:{\Bbb R}^{2}\times{\Bbb R}_{*}\rightarrow E$, where 
$\psi (x,y,i):=i(x,y)$. Due to the invariance of connection 1-forms under the right
action of ${\Bbb R}_{*}$, the reconstructed connection 1-form pulled back to 
${\Bbb R}^{2}\times{\Bbb R}_{*}$ can be written as:
\[
\hspace{35mm}
(C_{\psi}^{*}\omega )_{(x,y,z)}=A_{1}(x,y)dx+A_{2}(x,y)dy+\frac{A_{3}(x,y)}{z}dz
\hspace{35mm}(6.1)
\]
Furthermore,
\[
\frac{A_{3}(x,y)}{z}=C_{\psi}^{*}\omega\frac{\partial}{\partial z}(x,y,z)=C_{\psi
}^{*}\omega\frac{d}{di}(x,y,z+i)\mid_{i=0}=\phantom{iudfvyiaubfvuyabvfvuakbvbvbhbhj}
\]\[
=\omega\frac{d}{di}C_{\psi}(x,y,z+i)\mid_{i=0}=\omega\frac{d}{di}
\mbox{\boldmath $\{$}\widetilde{\psi}[(x,y)]\;\mbox{\boldmath $,$}\;
z+i\mbox{\boldmath $\}$}\mid_{i=0}=\phantom{iudfvyiaubfvuyabvfvuakbvbvbhbhj}
\]\[
=\mbox{\large\boldmath $($}
\mbox{\boldmath $\{$}\widetilde{\psi}[(x,y)]\;\mbox{\boldmath $,$}\;
z\mbox{\boldmath $\}$}^{-1}\mbox{\large\boldmath $)$}
_{*}\frac{d}{di}\mbox{\boldmath $\{$}
\widetilde{\psi}[(x,y)]\;\mbox{\boldmath $,$}\; z+i\mbox{\boldmath $\}$}\mid_{i=0
}=\frac{d}{di}\frac{z+i}{z}\mid_{i=0}=\frac{1}{z},\phantom{hjibfvbsyfgdbbsbhjs}
\]
i.e. $\! A_{3}(x,y)\equiv 1$. Now, let $\sigma :{\Bbb R}^{2}\rightarrow
{\Bbb R}^{2}\times{\Bbb R}_{*}$ be a global section defined by $\sigma 
(x,y)=(x,y,1)$. Using (6.1) and then (5.2), we obtain:
\[
A_{1}(x_{0},y_{0})=\sigma^{*}C_{\psi}^{*}\omega\frac{\partial}{\partial 
x}(x_{0},y_{0})=(C_{\psi}\circ\sigma)^{*}\omega\frac{\partial}{\partial 
x}(x_{0},y_{0})=\phantom{jhkdfivbdfyigbiyfuegjahkdbvhbvhjafbsvjahbv}
\]\[
=\frac{d}{dx}H\mbox{\boldmath $($}\widetilde{\psi}[(x,y_{0})]^{-1}\circ
T_{x,(x_{0},y_{0})}\circ\widetilde{\psi}[(x_{0},y_{0})]\mbox{\boldmath $)$}\mid
_{x=x_{0}}=\frac{d}{dx}exp\frac{y_{0}}{2}(x-x_{0})\mid
_{x=x_{0}}=\frac{y_{0}}{2}\phantom{djahfvghjagfdbvudfvjavfkvjknfjkdfjkn}
\vspace{3 mm}
\]

{\parindent=0pt
Similarly,
\[
A_{2}(x_{0},y_{0})=\frac{d}{dy}H\mbox{\boldmath $($}\widetilde{\psi}[(x_{0},y)]^{-1
}\circ
T_{y,(x_{0},y_{0})}\circ\widetilde{\psi}[(x_{0},y_{0})]\mbox{\boldmath $)$}\mid
_{y=y_{0}}
=\frac{d}{dy}exp\frac{-x_{0}}{2}(y-y_{0})\mid
_{y=y_{0}}=\frac{-x_{0}}{2}\]

Hence $(C_{\psi}^{*}\omega )_{(x,y,z)}=\frac{1}{2}ydx-\frac{1}{2}xdy+z^{-1}dz$.}\\

{\bf Acknowledgments:}
I am indebted to Andrzej Borowiec and Joseph Wolf for discussions concerning this 
article as well as for their encouragement. I also would like to thank Arkadiusz
Jadczyk for drawing my attention to the subject.

{\parindent=0pt
For more references and a discussion concerning the development of the subject
see \cite{b}.}

\end{document}